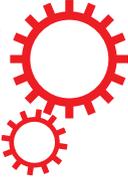

# Three representations of the Ising model

Joost Kruis[1] & Gunter Maris[1,2]

Statistical models that analyse (pairwise) relations between variables encompass assumptions about the underlying mechanism that generated the associations in the observed data. In the present paper we demonstrate that three Ising model representations exist that, although each proposes a distinct theoretical explanation for the observed associations, are mathematically equivalent. This equivalence allows the researcher to interpret the results of one model in three different ways. We illustrate the ramifications of this by discussing concepts that are conceived as problematic in their traditional explanation, yet when interpreted in the context of another explanation make immediate sense.

Scientific advances can be achieved by two types of theories: those that simply seek to identify correlations between observable events without regard to linking mechanisms; and those that specify the mechanisms governing the relations between observable events (Bandura, p. 21).[1].

Examining the structure of observed associations between measured variables is an integral part in many branches of science. At face value, associations (or their quantification in the form of correlations) inform about a possible relation between two variables, yet contain no information about the nature and directions of these relations. Making causal inferences from associations requires the specification of a mechanism that explains the emergence of the associations[2,3]. By constructing an explanatory model to account for associations in the data, that has testable consequences at the level of the joint distribution of variables, it is possible to test the adequacy of the model against the data. When the model is deemed sufficiently adequate with respect to the data, this is often perceived as justification for the proposed causal interpretation[4].

We can discern (at least) three general frameworks, each representing a different mechanism to explain the emergence of associations between variables, with their own collection of corresponding explanatory models. These frameworks, and their corresponding statistical models, all originate from different disciplines and have received considerable attention in diverse fields such as, physics, mathematics, statistical mechanics, causality, biology, epidemiology, and social sciences. Although both authors originate from psychology, and primarily illustrate their findings with examples from this field, the frameworks and models discussed in this paper clearly transcend the domain of psychology, and as such have a multidisciplinary relevance. In the current paper we refer to these frameworks as, respectively, the common cause-, reciprocal affect-, and common effect framework.

The *common cause framework* explains the observed associations through a latent (unobserved) variable acting as a common cause with respect to the manifest (observed) variables[5]. Causal models that propose a common cause mechanism as generating the associations between manifest variables, are also known as reflective models[6,7]; the manifest variables are indicators of the latent variable and reflect its current state. In the statistical literature, models in this framework are therefore often referred to as latent variable models. Latent variable models have proven to be extremely successful at fitting observed multivariate distributions, at the same time their theoretical and philosophical underpinning remains problematic. Latent variables are both a powerful and controversial concept, in psychology for example, the idea of psychological constructs as intelligence[8,9] and personality[10] being latent variables has been the subject of many intense debate. In particular about the question whether one should take a realist interpretation of latent variables, that is the latent variable signifying a real but hidden entity, to justify the use of latent variable analysis[11,12]. An important reason for this is that a latent cause is never observed, and similar to physics around the turn of the 20th century, there was need for…

Popper, p. 211[13]. … an epistemological programme: to rid the theory of 'unobservables', that is, of magnitudes inaccessible to experimental observation; to rid it, one might say, of metaphysical elements.

[1]University of Amsterdam, Department of Psychology, Amsterdam, The Netherlands. [2]Cito, Psychometric Research Center, Arnhem, The Netherlands. Correspondence and requests for materials should be addressed to J.K. (email: j.kruis@uva.nl)

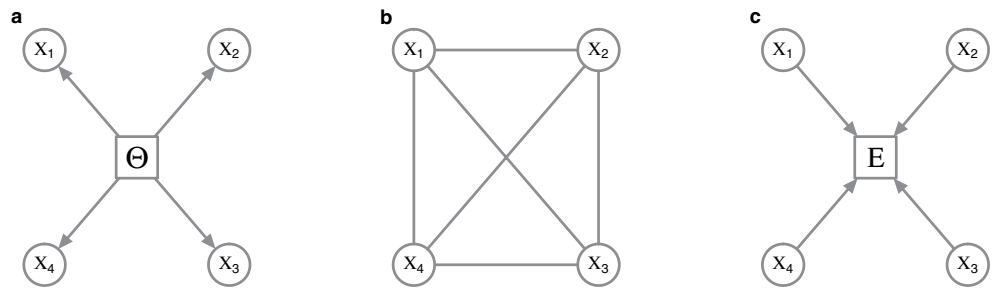

**Figure 1. Three frameworks for explaining observed associations in the parametrisation of their prototypical statistical models.** (**a**) The Rasch model from the common cause framework as a DAG. (**b**) The Curie-Weiss model from the reciprocal affect framework as an undirected graph. (**c**) The collider selection bias model from the common effect framework as a DAG.

It is in the *reciprocal affect framework* that we find such a programme without 'unobservables'. In this framework the relations between variables are represented as is, in that, observed associations between manifest variables are explained as a consequence of mutualistic relations between these variables[14]. The idea that variables are associated as a consequence of reciprocal affect has been formalised in the field of network analysis[15], and has been studied extensively in diverse field of science such as mathematics, physics, and biology[16–19]. The Ising model is a suitable example of a statistical model in this framework, as it captures all main effects and pairwise interactions between variables[20,21]. While originally a model of ferromagnetism from statistical physics, in the last decade the Ising model has been adopted within the social sciences, where the network perspective has been gaining much popularity[22–25]. As an alternative for the latent variable perspective, the network approach has lead to valuable new insights about, for example, psychopathology and the aforementioned concept of intelligence[26].

In the, final, *common effect framework*, observed associations between manifest variables are explained as arising from (unknowingly) conditioning on a common effect of these variables; the manifest variables are marginally independent with respect to each other, yet their collective state leads to the occurrence (or absence) of the effect[27,28]. Variables can act as the collective cause towards an effect in (at least) two ways, either as the separate indicators of an artificial compound score (e.g. Socio-Economic-Status, SAT)[29], or as determinants of a naturally occurring phenomenon, such as agents culminating into the outbreak of an epidemic. In the statistical literature, the term collider variable models is used for this framework, as the collective state of the independent variables collides into the effect[4,30]. Because of the independence, one would naturally expect not to find any associations between these variables. However, from the literature on causality it is known that conditioning on a collider variable introduces (spurious) correlations among the variables functioning as the collective cause. This phenomenon is known as endogenous selection bias and will results in the observation of associations between the manifest variables[30–34].

It is clear that each of these frameworks proposes a radically different explanation for the emergence of associations between a set of manifest variables. In this paper we argue that these differences only exist with respect to the theoretical interpretation of these frameworks. Specifically, we demonstrate that the prototypical statistical models for binary data in each framework are mathematically equivalent, and that this equivalence extends to more realistic models that capture all main effects and pairwise interactions between the observed variables. Through this we obtain three, statistically equivalent, representations of the Ising model that each explain the occurrence of associations between binary variables by a theoretically very distinct mechanism.

## Results

**Prototypical models.** To enhance the readability of this section we start by introducing the variables that return in all discussed models, and clarify the mathematical notation used in the text and equations for the distribution functions. We will denote random variables with capital letters and possible realisations of random variables with lower case letters. We represent vectors with bold-faced letters, and use boldfaced capital letters to indicate matrices for parameters. Manifest variables are denoted with Roman letters, whereas we use Greek letters to indicate latent variables and parameters that need be estimated.

In the context of the paper we are primarily interested in the vector $\mathbf{X} = [X_1, X_2, \ldots, X_N]$, consisting of $N$ binary random variables that can take $+1$ and $-1$ as values, as we look to the mechanism by which the three frameworks explain the observed associations between the realisations of this vector denoted by $\mathbf{x} = [x_1, x_2, \ldots, x_n]$. Furthermore, each of the models we discuss includes a vector containing the main effect parameters $\boldsymbol{\delta} = [\delta_1, \delta_2, \ldots, \delta_n]$, consisting of $N$ numbers in $\mathbb{R}$. Except for equation (1) which we write out in full, we use $\sum_i$ and $\prod_i$ to denote respectively $\sum_{i=1}^n$ and $\prod_{i=1}^n$ for the remainder of the equations. Additionally, we use $p(\mathbf{x})$ to denote $p(\mathbf{X} = \mathbf{x})$, which extends to all variables in both conditional, and joint probability distributions, such that we can read it as the probability of observing some realisation of the random vector $\mathbf{X}$, optionally, conditional on, or together with, the realisation of some other variable.

We consider the Rasch model[35], an Item Response Theory (IRT) model from the field of psychometrics, as the prototypical model for binary data in the common cause framework. Historically, the Rasch model has been developed for modelling the responses of persons to binary scored items on a test. The model is graphically represented in Fig. 1(a) as a Directed Acyclic Graph (DAG)[36], where the latent random variable $\Theta$ acts as the common cause of the manifest random variables **X**. The Rasch model is characterised by the following distribution for the manifest variables (**X**), conditional on the latent variable ($\Theta \in \mathbb{R}$):

$$p(\mathbf{X} = \mathbf{x}|\Theta = \theta) = \prod_i \frac{\exp(x_i[\theta + \delta_i])}{\exp(+[\theta + \delta_i]) + \exp(-[\theta + \delta_i])} \quad (1)$$

The marginal probabilities for **X** can be obtained by endowing $\Theta$ with a distribution as shown in equation (2). While this gives us an expression for the manifest probabilities of the Rasch model, for almost all choices for the distribution of $f(\theta)$, this expression becomes computationally intractable.

$$p(\mathbf{x}) = \int_{\mathbb{R}} \prod_i \frac{\exp(x_i[\theta + \delta_i])}{\exp(+[\theta + \delta_i]) + \exp(-[\theta + \delta_i])} f(\theta) \, d\theta \quad (2)$$

In the traditional interpretation of the Rasch model, $x_i$ indicates whether the response to item *i* is correct ($x_i = +1$) or incorrect ($x_i = -1$). In this context, the continuous random variable ($\Theta$) represents the latent ability being examined by the set of items (**X**). The vector **δ** contains the item main effects, where $\delta_i$ represents the easiness of item *i*, such that $-\delta_i$ represents the difficulty of item *i* with respect to the measured ability. The response of an individual on item *i* is a trade-off between the ability of the person ($\theta$) and the item difficulty ($-\delta_i$). When the ability of the person is greater than the difficulty of the item, the probability for a correct response will be higher than for an incorrect response ($\theta > -\delta_i \Rightarrow p(x_i = 1) > p(x_i = -1)$), if the ability of the person is lower than the item difficulty the reverse holds ($\theta < -\delta_i \Rightarrow p(x_i = 1) < p(x_i = -1)$). As such, persons with a greater ability will always have a higher probability of giving a correct response, and persons always has a higher probability for a correct response on an easy item than on a more difficult item. A key property of the Rasch model is that of *local independence*, which entails that only variation in $\Theta$ determines the probability for a response on an item. That is, conditionally on the state of the latent variable all manifest variables are independent, such that marginally (with respect to the latent variable) they are dependent. Consequently, any observed associations between the manifest variables can be traced back to the influence of the latent variable. It is the latent variable that causes the associations on the manifest variables, which is why the Rasch model falls within the common cause framework.

For the reciprocal affect framework we examine the Curie-Weiss model from statistical physics[37–39], originally used to model the state of a set of magnetic moments, for which the thermodynamical properties correspond to that of the classical Curie-Weiss theory of magnetism, and where the pairwise interactions between the magnetic moments are replaced by the mean magnetisation. Graphically, the Curie-Weiss model can be represented as an undirected graph wherein the manifest variables (**X**), representing the set of magnetic moments, are fully connected with each other, and all connection are of equal strength, as illustrated in Fig. 1(b). The distribution of the manifest variables (**X**) in the Curie-Weiss model is given by:

$$p(\mathbf{x}) = \frac{1}{Z} \exp\left(\sum_i x_i \delta_i + \frac{1}{2}\left[\sum_i x_i\right]^2\right) \quad (3)$$

In the conventional interpretation of the Curie-Weiss model, $x_i$ indicates that the magnetic spin of moment *i* is upward ($x_i = +1$) or downward ($x_i = -1$), whereas the main effect for each moment ($\delta_i$) indicates the natural preference of moment *i* to be in an upward ($\delta_i > 0$) or downward ($\delta_i < 0$) spin position, due to the external magnetic field not present in **X**. In equation (3) we use $Z$ to represents the normalising constant, in thermodynamical systems often referred to as the partition function, that makes the distribution sum to one. In the Curie-Weiss model, this partition function sums over all $2^N$ possible configurations of the vector **X**, which we denote in this paper as $\sum_{\mathbf{x}}$, and is given by the following expression:

$$Z = \sum_{\mathbf{x}} \exp\left(\sum_i x_i \delta_i + \frac{1}{2}\left[\sum_i x_i\right]^2\right) \quad (4)$$

As the pairwise interactions between the magnetic moments are replaced by the mean magnetisation, all interactions are captured by the squared sum of the set of moments in the exponential of the Curie-Weiss distribution. By averaging over individual interactions between magnetic moments, the Curie-Weiss model is the simplest non-trivial model that exhibits a phase transition in statistical mechanics. However, because the model violates fundamental principles of statistical physics, and its predictions are only partially verified by experiments, it is considered as being mainly of theoretical interest[37]. Nonetheless, due to its simplicity, the Curie-Weiss model has been useful in understanding the dynamics of equivalent phenomena in more realistic systems, such as the Ising model[39]. Still, it is clear that as the observed associations between magnetic moments are presumed to emerge due to the magnetic interaction between these moments themselves, the Curie-Weiss model falls within the reciprocal affect framework.

Whereas in the Rasch model, from the common cause framework, associations between manifest variables are explained by the latent variable $\Theta$, in the Curie-Weiss model, from the reciprocal affect framework, these associations are captured in the squared sum of the set of moments in the exponential of the distribution. The key ingredient for establishing the connection between these two models has been known for a long time, and has also been rediscovered quite a few times in quite diverse fields of science[37,40–45]. It was in his Brandeis lecture that Mark Kac[37] established the relation between the Curie-Weiss model and the Rasch model through an ingenuous use of the following Gaussian integral:

$$\exp(a^2) = \int_{\mathbb{R}} \frac{\exp(2a\theta - \theta^2)}{\sqrt{\pi}} d\theta \qquad (5)$$

What Kac realized is that whenever you see the exponential of a square, you can replace it with the right hand side integral from equation (5). In the Methods section we demonstrate that applying this Gaussian identity to the Curie-Weiss distribution from equation (3) linearises the squared sum in the exponential, and introduces a random variable $\Theta$, such that we obtain a latent variable representation of the Curie-Weiss model. We then show that because the square in the exponential is gone, we can rewrite the expression for the latent variable representation of the Curie-Weiss model such that, both the marginal distribution of the manifest variables, and that of the manifest variables conditional on $\theta$, are identical to that of the Rasch model from equation (1) and equation (2). Having established the relation between the prototypical models from the common cause, and reciprocal affect framework we turn to our third framework.

For the common effect framework we consider **X** as a set of independent random variables, which we will collectively call the cause, together with a single dependent binary random variable ($E$), which we will call the effect. Their joint distribution, given in equation (6), is a collider structure and can be graphically represented in a DAG as illustrated in Fig. 1(c).

$$p(\mathbf{x}, e) = \sum_i \frac{\exp(x_i \delta_i)}{\exp(+\delta_i) + \exp(-\delta_i)} \left( \frac{\exp(\frac{1}{2}[\sum_i x_i]^2)}{\sup_{\mathbf{x}} \exp(\frac{1}{2}[\sum_i x_i]^2)} \right)^e$$
$$\times \left( 1 - \frac{\exp(\frac{1}{2}[\sum_i x_i]^2)}{\sup_{\mathbf{x}} \exp(\frac{1}{2}[\sum_i x_i]^2)} \right)^{1-e} \qquad (6)$$

In this collider distribution, $x_i$ indicates if cause $i$ is active ($x_i = +1$) or inactive ($x_i = -1$), whereas ($e$) indicates whether the effect was present ($e=1$) or absent ($e=0$) at that time. The main effect for each cause ($\delta_i$) denotes the natural predisposition for cause $i$ to be active ($\delta_i > 0$) or inactive ($\delta_i < 0$) at any given time. As mentioned, and shown by equation (9) in the Methods section, are the individual causes independent of each other in the marginal distribution of **X**. As a consequence when we marginalise **X** with respect to $E$, the causes will not show any associations among each other. From the literature on causality it is however known that selection with respect to a common effect variable will introduce (spurious) correlations among the causes. That is, by using only observations of **x** where the common effect is present ($e=1$), this set of observations will show a pattern of associations among the causes. This is known as endogenous selection bias with respect to a collider variable, and can be mathematically represented in the distribution of the causes conditionally on the effect. In the Methods section we demonstrate that when we apply this selection bias mechanism to the collider structure from equation (6), the distribution of the collective cause (**X**) conditionally on the effect, exactly gives the Curie-Weiss model from statistical physics, and hence, the Rasch model.

**Realistic models.** Having studied the three statistical explanations in their simplest non-trivial form, we conclude that, although their theoretical interpretation is radically different, the three models are mathematically indistinguishable. Still, the simplicity of the prototypical models for each framework also makes them often unrealistic with respect to the observed reality. Specifically, the Rasch model and simple collider model only allow for main effects between the observations, and do not consider possible pairwise interactions. The Curie-Weiss model does allow some crude form of interaction, however, as the individual interactions between nearest neighbours are replaced by the mean interaction, one must make the (often) unrealistic assumption that all observations are interconnected with the same strength. Fortunately, we can swiftly generalise all three prototypical models to more realistic forms.

We start with the Ising model, of which the Curie-Weiss model is the simplest form, from the reciprocal affect framework. Like the Curie-Weiss model, the Ising model was originally introduced in statistical physics as a model for magnetism, with the same possible values and interpretation for **X** and its possible realisations. However, instead of only considering the mean magnetisation, the Ising model captures all pairwise interactions between the set of manifest variables (**X**). The distribution of the Ising model, where $\sum_{\langle i,j \rangle}$ is the sum over all distinct pairs of magnetic moments, is commonly written as follows:

$$p(\mathbf{x}) = \frac{1}{Z} \exp\left( \sum_i x_i \delta_i + \sum_{\langle i,j \rangle} x_i x_j \sigma_{ij} \right) \qquad (7)$$

The pairwise interactions are represented in the Ising model distribution by the symmetric $N \times N$ connectivity matrix $\Sigma$ in $\mathbb{R}$. In this connectivity matrix, $\sigma_{ij}$ modulates the reciprocal affect relation between $x_i$ and $x_j$, indicating if moments $i$ and $j$ prefer to be in identical ($\sigma_{ij} > 0$), or opposing ($\sigma_{ij} < 0$) spin positions, wherein the higher the absolute value of $\sigma_{ij}$, the stronger this preference. Under the condition that all off-diagonal entries of $\Sigma$ are equal, the Ising model reduces to the prototypical Curie-Weiss model. Because the diagonal values of the connectivity matrix in the Ising model are arbitrary, i.e., the probability of **X** is independent of these values, we can choose the values for the diagonal in such a way that the connectivity matrix becomes positive (semi) definite. As a consequence the eigenvalue decomposition of the matrix will also be non-negative. As clarified in the Methods section, by applying this transformation to the Ising model distribution from equation (7) we obtain an eigenvalue representation of the Ising model:

$$p(\mathbf{x}) = \frac{1}{Z} \exp\left(\sum_i x_i \delta_i + \sum_r \frac{1}{2} \lambda_r \left[\sum_i q_{ir} x_i\right]^2\right) \tag{8}$$

where $\lambda_r$ is the $r^{th}$ non-negative eigenvalue of the vector $\boldsymbol{\Lambda} = [\lambda_1, \lambda_2, \ldots, \lambda_N]$, and $q_{ir}$ the value of the $i^{th}$ row and $r^{th}$ column of the $N \times N$ eigenvector matrix $\mathbf{Q}$. In the equation above $\sum_r$ should be read as $\sum_{r=1}^{n}$, we continue this practice in the notation of the applicable equations in the Methods section. In the Methods section we demonstrate how this eigenvalue representation allows us to connect the Ising model from the reciprocal affect framework to the more realistic models in both the other frameworks. First, by applying the Gaussian identity from Kac to the squared sum in the exponent for each eigenvalue in equation (8), we obtain a latent variable representation of the Ising model[46], with as many latent dimensions as there are non-zero eigenvalues. This latent variable representation of the Ising model is then shown to be the multidimensional IRT model[47] from the common cause framework, of which the Rasch model is the simplest instance, but allows for more than one latent variable to explain the observed associations between the manifest variables. Similarly, we can introduce (independent) effect variables for each eigenvalue in equation (8), such that we obtain a collider representation of the Ising model where endogenous selection bias has taken place. We then show that this distribution is a version of the common effect model as seen in equation (6), that is extended such that the collective cause can collide into more then one common effect.

## Discussion

We have shown that the mathematical equivalence of the simple prototypical models from the common cause, reciprocal affect, and common effect framework, extends to the more realistic counterparts of these models. That is, there exist three, statistically indistinguishable, representations of the Ising model that explain observed associations either through marginalisation with respect to latent variables, through reciprocal affect between variables, or through conditioning on common effect variables. We therefore argue that these are not three different models, but just one model for which three distinct theoretical interpretations have been developed in different fields of science. Consequently, any set of associations between variables that is sufficiently described by a model in one framework, can be explained as emerging from the mechanism represented by any of the three theoretical frameworks. We illustrate the implications of this by considering one of the most controversial topics in the common cause framework, differential item functioning (DIF)[48], and discuss it in the context of the three possible interpretations.

In it's traditional (common cause) framework DIF indicates that, conditional on the level of the latent variable, the probability for some response is dependent on group membership. For items that exhibit DIF it is not only variation in Θ that determines the probability for a response on an item. From a common cause perspective the occurrence of DIF is a violation of local independence, and as such measurement invariance. In the context of ability testing DIF is often perceived as indication of item bias[49]. As a fictitious example, consider the situation where on certain items from the Revised NEO Personality Inventory (NEO PI-R)[50], that intents to measure the Big Five personality traits[51], we find that for a group of subjects with the same latent trait score, those that listed their occupation as being a manager always have a higher probability of giving a correct response on these items, compared to subjects that have no occupation as a manager.

Needless to say, in this context items that exhibit DIF are seen as bad because they pose a problem for both the reliability and validity of a test. From a reciprocal affect perspective, identification of DIF would exhibit itself in the form of differences in the estimated pairwise associations between items depending on group membership. As such the appearance of DIF in the NEO PI-R example would also be viewed as troublesome. However, in contrast to the common cause framework, the appearance of DIF in a network model is at least informative in that our model might be incomplete, i.e., the network is missing a node. The interpretation of DIF in a common effect framework is best understood, in the context of the current example, by considering the answer to the question: *What causes people to obtain a managerial position as occupation?* It is safe to say that in most cases a persons personality is an important factor in this process. In other words, people that are selected to become manager, get this position because they posses a certain set of personality traits associated with being a successful manager[52]. As such, the items in the NEO PI-R that show DIF in this case measure those personality traits that are most sought after in managers. More broadly in the context of the common effect framework, the occurrence of DIF indicates how well an item predicts differences in the effect. In contrast to the disruptive interpretation of DIF in both the common cause and reciprocal affect frameworks, the occurrence of DIF within the context of the common effect framework is actually both sensible and informative.

The previous example clearly demonstrates how fundamental concepts, that are firmly established in their traditional framework as being problematic, can be perceived as neutral and informative or even desirable in another context. Having multiple possible interpretations for the same model allows for more plausible explanations when it comes to the theoretical concepts and the causal inferences we obtain from the measurement model applied to our data. For example in the context of psychopathology, depression has been habitually being treated as a common cause variable for which its symptoms are the interchangeable indicators. Measures of these symptoms with the popular Beck Depression Inventory[53] have shown to fit a latent variable model with one underlying general depression factor and three highly inter-correlated sub-factors, or a two-factor solution well[54]. However, in a common cause framework depression symptoms, such as sleep problems, loss of energy, and trouble concentrating, are assumed independent of one another, as they are purely caused by the latent variable interpreted as depression. More recently it has been shown that a network model can also give an accurate description of data on depression symptoms[26]. The reciprocal affect representation of depression, where symptoms can directly influence each other, is as an explanation more in line with our perceived reality. Furthermore, the historical success of theoretically very implausible models, such as the latent variable model, can thus in retrospect, arguably be explained by the equivalence of these three models.

Being able to interpret the outcome of an applied measurement model from theoretically very distinct perspectives, instead of only the perspective as traditionally assumed by the model, is great progress, as it allows for novel explanations that might be a better reflection of our perceived reality. Furthermore, in their different fields of application different aspects of these models have been studied and different methodology has been developed. Through their connection much of these developments become available to all fields of application.

## Methods

In this section we clarify the mathematics involved in connecting the simple prototypical models, as well as the more realistic Ising model representations, for the three different frameworks. In the first proof we demonstrate the equivalence between the simple collider, Curie-Weiss and Rasch models, the prototypical (yet unrealistic) models for respectively the common effect, reciprocal affect, and common cause explanation for observed associations between a set of binary variables.

### Proof for the equivalence of the simple prototypical models.    *Collider to Curie-Weiss.*    The simple collider model from the common effect framework is characterised by the following joint probability distribution $p(\mathbf{x}, e)$:

$$p(\mathbf{x}, e) = \prod_i \frac{\exp(x_i \delta_i)}{\exp(+\delta_i) + \exp(-\delta_i)} \left( \frac{\exp\left(\frac{1}{2}[\sum_i x_i]^2\right)}{\sup_{\mathbf{x}} \exp\left(\frac{1}{2}[\sum_i x_i]^2\right)} \right)^e$$

$$\times \left( 1 - \frac{\exp\left(\frac{1}{2}[\sum_i x_i]^2\right)}{\sup_{\mathbf{x}} \exp\left(\frac{1}{2}[\sum_i x_i]^2\right)} \right)^{1-e}$$

(9)

In order to connect this collider model to the Curie-Weiss model we introduce endogenous selection bias on the set of manifest variables forming the collective cause, by conditioning on the effect being present. This is mathematically presented as the conditional distribution $p(\mathbf{x}|e=1)$, proportional to the product of the marginal distribution for the cause $p(\mathbf{x})$, and the probability of observing the effect given the cause $p(e=1|\mathbf{x})$, defined by:

$$p(\mathbf{x}|e=1) \propto p(\mathbf{x})p(e=1|\mathbf{x}) = \prod_i \frac{\exp(x_i \delta_i)}{\exp(+\delta_i) + \exp(-\delta_i)} \frac{\exp\left(\frac{1}{2}[\sum_i x_i]^2\right)}{\sup_{\mathbf{x}} \exp\left(\frac{1}{2}[\sum_i x_i]^2\right)}$$

(10)

We can simplify the expression for $p(\mathbf{x}|e=1)$ by recognising that the product of exponentials in the numerator can be rewritten as a sum within the exponential. Furthermore, the denominator of the expression is only dependent on the sum of $\mathbf{X}$, and thus independent of the specific pattern that the realisation of $\mathbf{X}$ takes. As a consequence $p(\mathbf{x}|e=1)$ is only proportional to the numerator of equation (10), such that we can write:

$$p(\mathbf{x}|e=1) \propto \exp\left(\sum_i x_i \delta_i + \frac{1}{2}\left[\sum_i x_i\right]^2\right)$$

(11)

In order to obtain a valid probability density function we have to add the appropriate normalising constant that makes the probabilities sum to one again. In this case this translates to dividing the expression in equation (11) for a certain realisation of $\mathbf{X}$ by the sum of this expression for all possible configurations of $\mathbf{X}$:

$$p(\mathbf{x}|e=1) = \frac{\exp\left(\sum_i x_i \delta_i + \frac{1}{2}[\sum_i x_i]^2\right)}{\sum_{\mathbf{x}} \exp\left(\sum_i x_i \delta_i + \frac{1}{2}[\sum_i x_i]^2\right)}$$

(12)

It can quickly be verified that the resulting expression in equation (12) is identical to the distribution for the Curie-Weiss model introduced in equation (3), with the same normalising constant as given in equation (4). Thus proofing that, conditional on the effect being present, the distribution of the collective cause in the collider model is equivalent to the distribution of a set of directly interacting magnetic moments in the Curie-Weiss model.

*Curie-Weiss to Rasch.*    Next, we will connect the Curie-Weiss model from statistical physics to the Rasch model from psychometrics. We start from the distribution function of the Curie-Weiss model, where we use $Z$ to denote the appropriate normalising constant:

$$p(\mathbf{x}) = \frac{\exp\left(\sum_i x_i \delta_i + \frac{1}{2}[\sum_i x_i]^2\right)}{Z}$$

(13)

Next we use Kac's Gaussian identity from equation (5) to linearise the quadratic sum in the exponential of the Curie-Weiss distribution, to that end let $a^2 = \frac{1}{2}[\sum_i x_i]^2$, so we can rewrite it in the following way:

$$\exp\left(\frac{1}{2}[\sum_i x_i]^2\right) = \int_{\mathbb{R}} \frac{\exp\left(2\frac{1}{2}[\sum_i x_i]\theta - \theta^2\right)}{\sqrt{\pi}} d\theta \qquad (14)$$

By incorporating this transformation we obtain a latent variable representation of the Curie-Weiss model

$$p(\mathbf{x}) = \int_{\mathbb{R}} \frac{\exp(\sum_i x_i \delta_i + [\sum_i x_i]\theta - \theta^2)}{Z\sqrt{\pi}} d\theta \qquad (15)$$

Which we can simplify further by merging the two sums in the exponential:

$$p(\mathbf{x}) = \int_{\mathbb{R}} \frac{\exp(\sum_i x_i [\theta + \delta_i] - \theta^2)}{Z\sqrt{\pi}} d\theta \qquad (16)$$

Where we will use $Z^* = Z\sqrt{\pi}$ to denote the appropriate normalising constant. For the next step towards our goal we multiply both the numerator and denominator of equation (16) by $\prod_i [\exp(+[\theta + \delta_i]) + \exp(-[\theta + \delta_i])]$, such that we obtain the equivalent expression:

$$p(\mathbf{x}) = \int_{\mathbb{R}} \frac{\exp(\sum_i x_i [\theta + \delta_i] - \theta^2)}{Z^*} \frac{\prod_i [\exp(+[\theta + \delta_i]) + \exp(-[\theta + \delta_i])]}{\prod_i [\exp(+[\theta + \delta_i]) + \exp(-[\theta + \delta_i])]} d\theta \qquad (17)$$

Next we rearrange the expression in equation (17) by switching the denominators of both factors, taking the sum in the first numerator out of the exponential so it becomes a product, and transferring $\exp(-\theta^2)$ out the numerator of the first factor, and into the numerator of the second factor:

$$\begin{aligned} p(\mathbf{x}) &= \int_{\mathbb{R}} \prod_i \frac{\exp(x_i[\theta + \delta_i])}{\exp(+[\theta + \delta_i]) + \exp(-[\theta + \delta_i])} \\ &\quad \times \frac{\prod_i [\exp(+[\theta + \delta_i]) + \exp(-[\theta + \delta_i])]\exp(-\theta^2)}{Z^*} d\theta \end{aligned} \qquad (18)$$

The resulting expression can be recognised as $p(\mathbf{x}) = \int_{\mathbb{R}} p(\mathbf{x}|\theta) f(\theta) d\theta$, the marginal probability for some realisation of **X** where the latent variable $\Theta$ is integrated out. Let us denote the second factor of the expression in equation (18) as the distribution of the latent variable ($f(\theta)$), which gives us:

$$p(\mathbf{x}) = \int_{\mathbb{R}} \prod_i \frac{\exp(x_i[\theta + \delta_i])}{\exp(+[\theta + \delta_i]) + \exp(-[\theta + \delta_i])} f(\theta) d\theta \qquad (19)$$

Such that the distribution of the set of binary random variables (**X**), conditionally on the latent variable ($\Theta$), is:

$$p(\mathbf{x}|\theta) = \prod_i \frac{\exp(x_i[\theta + \delta_i])}{\exp(+[\theta + \delta_i]) + \exp(-[\theta + \delta_i])} \qquad (20)$$

Again, it is readily seen that the resulting latent variable expression of the Curie-Weiss model in equation (20) is identical to the distribution of the Rasch model from equation (1). This completes our first proof in which we demonstrated that by conditioning on the effect in the collider model from the common effect framework, the distribution of set of binary random variables (**X**) is equivalent to that of the Curie-Weiss model from the reciprocal affect framework. Furthermore, when we linearise the quadratic sum in the exponential of the Curie-Weiss model, we obtain a latent variable representation of this model where the distribution of the manifest random variables (**X**) given the latent variable ($\Theta$) is equivalent to that of the Rasch model from the common cause framework. Consequently, given the equivalence of the collider model and the Curie-Weiss model, and that of the Curie-Weiss model and the Rasch model, we can conclude that the collider model and the Rasch model are also equivalent.

In the next proof we demonstrate that this equivalence relation between the three frameworks extends to the more realistic models of these frameworks, as those allow pairwise interactions between the random variables in the set **X**. We start with the conventional representation of the full Ising model from the reciprocal affect framework and rewrite this into an equivalent eigenvalue representation. Subsequently we connect this to both a latent variable representation equivalent to the multidimensional IRT model from the common cause framework, and a collider representation from the common effect framework.

**Three representations of the Ising model.** *Conventional to Eigenvalue representation.* The distribution of the Ising model is commonly written as follows:

$$p(\mathbf{x}) = \frac{1}{Z} \exp\left(\sum_i x_i \delta_i + \sum_{\langle i,j \rangle} x_i x_j \sigma_{ij}\right) \qquad (21)$$

Where the partition function Z, that makes the distribution sum to one, is given by:

$$Z = \sum_{\mathbf{x}} \exp\left(\sum_i x_i \delta_i + \sum_{\langle i,j \rangle} x_i x_j \sigma_{ij}\right) \tag{22}$$

In order to connect the Ising model from the reciprocal affect framework to the models from both other frameworks, we first have to rewrite it into matrix notation such that we can obtain the eigenvalue representation of the Ising model. To that end we first rewrite the sum over all distinct $i, j$ pairs in the exponent, as a function of the sum over $i$ and the sum over $j$:

$$p(\mathbf{x}) = \frac{1}{Z} \exp\left(\sum_i x_i \delta_i + \frac{1}{2}\sum_{i=1}^{n}\sum_{j=1}^{n} x_i x_j \sigma_{ij}\right) \tag{23}$$

Such that we may rewrite the Ising model in matrix notation:

$$p(\mathbf{x}) = \frac{1}{Z} \exp\left(\mathbf{x}^T \boldsymbol{\delta} + \frac{1}{2}\mathbf{x}^T \boldsymbol{\Sigma} \mathbf{x}\right) \tag{24}$$

All parameters, except for entries on the diagonal of the connectivity matrix, are identifiable from the data. However, as $x_i x_j = 1$ when $i = j$ any diagonal entry for the connectivity matrix will be cancelled out by the partition function. With the observation that the diagonal values of $\boldsymbol{\Sigma}$ are thus arbitrary (i.e., do not change the probabilities), we can shift them in such a way that the connectivity matrix $\boldsymbol{\Sigma}$ becomes positive (semi) definite, and hence its eigenvalue decomposition non-negative. This allows for the transformation $\boldsymbol{\Sigma} + c\mathbf{I} = \mathbf{Q}\boldsymbol{\Lambda}\mathbf{Q}^T$, where $c$ contains the chosen values for the diagonal of the connectivity matrix, that when implemented gives:

$$p(\mathbf{x}) = \frac{1}{Z} \exp\left(\mathbf{x}^T \boldsymbol{\delta} + \frac{1}{2}\mathbf{x}^T \mathbf{Q} \boldsymbol{\Lambda} \mathbf{Q}^T \mathbf{x}\right) \tag{25}$$

By taking the expression out of its matrix notation we obtain the eigenvalue representation of the Ising model:

$$p(\mathbf{x}) = \frac{1}{Z} \exp\left(\sum_i x_i \delta_i + \sum_r \frac{1}{2}\lambda_r \left[\sum_i q_{ir} x_i\right]^2\right) \tag{26}$$

*Eigenvalue to Latent Variable representation.* We obtain a latent variable representation of the Ising model by applying Kac's Gaussian identity to the squared sum in the exponent of equation (26). To that end let $a^2$ be $\frac{1}{2}\lambda_r[\sum_i q_{ir} x_i]^2$ for each of the $N$ eigenvalues in the Ising model, and replace this with the right hand side integral from equation (5):

$$\exp\left(\frac{1}{2}\lambda_r\left[\sum_i q_{ir} x_i\right]^2\right) = \int_{\mathbb{R}} \frac{1}{\sqrt{\pi}} \exp\left(2\frac{1}{2}\lambda_r\left[\sum_i q_{ir} x_i\right]\theta_r - \theta_r^2\right) d\theta_r \tag{27}$$

Incorporating this transformation into the Ising model, and letting $Z^\# = Z\sqrt{\pi}^N$, we get he latent variable representation of the Ising model, where the number of non-zero eigenvalues represents the number of latent dimensions in the model:

$$p(\mathbf{x}) = \int_{\mathbb{R}} \frac{1}{Z^\#} \exp\left(\sum_i x_i \delta_i + \sum_r \lambda_r\left[\sum_i q_{ir} x_i\right]\theta_r - \theta_r^2\right) d\boldsymbol{\theta} \tag{28}$$

In order to connect this latent variable representation to the multidimensional IRT model we the multiply we multiply both the numerator and denominator of equation (28) by $\sum_{\mathbf{x}} \exp(\sum_i x_i \delta_i + \sum_r \lambda_r[\sum_i q_{ir} x_i]\theta_r)$, such that we obtain the equivalent expression:

$$p(\mathbf{x}) = \int_{\mathbb{R}} \frac{\exp(\sum_i x_i \delta_i + \sum_r \lambda_r[\sum_i q_{ir} x_i]\theta_r - \theta_r^2)}{Z^\#} \\ \times \frac{\sum_{\mathbf{x}} \exp(\sum_i x_i \delta_i + \sum_r \lambda_r[\sum_i q_{ir} x_i]\theta_r)}{\sum_{\mathbf{x}} \exp(\sum_i x_i \delta_i + \sum_r \lambda_r[\sum_i q_{ir} x_i]\theta_r)} d\boldsymbol{\theta} \tag{29}$$

Next we rearange and simplify the expression in equation (29). To that end let us merge the sums over $x_i$ in the exponential, and denote $\lambda_r q_{ir}$ as $\alpha_{ir}$, where $\alpha_{ir}$ is the value of the $i^{th}$ row and $r^{th}$ column of the $N \times N$ matrix $\mathbf{A}$ in $\mathbb{R}$. We can then rewrite the sum over $r$ as a product of the vector $\boldsymbol{\alpha}_i^T$, containing the $i^{th}$ row of the matrix $\mathbf{A}$, and the vector $\boldsymbol{\theta}$. Furthermore, we switch the denominators of both factors and transfer $\exp(\sum_r \theta_r^2)$ out the numerator of the first factor, to the numerator of the second factor:

$$p(\mathbf{x}) = \int_{\mathbb{R}} \frac{\exp(\sum_i x_i[\delta_i + \boldsymbol{\alpha}_i^T \boldsymbol{\theta}])}{\sum_{\mathbf{x}} \exp(\sum_i x_i[\delta_i + \boldsymbol{\alpha}_i^T \boldsymbol{\theta}])} \frac{\sum_{\mathbf{x}} \exp(\sum_i x_i[\delta_i + \boldsymbol{\alpha}_i^T \boldsymbol{\theta}] - \sum_r \theta_r^2)}{Z^\#} d\boldsymbol{\theta} \qquad (30)$$

In the resulting expression we can again recognise a function for the marginal probability of **X** where all latent dimensions (**Θ**) are integrated out. Finally we take the sum over *i* out of the exponential such that we obtain:

$$p(\mathbf{x}) = \int_{\mathbb{R}} \prod_i \frac{\exp(x_i[\delta_i + \boldsymbol{\alpha}_i^T \boldsymbol{\theta}])}{\sum_{\mathbf{x}} \exp(x_i[\delta_i + \boldsymbol{\alpha}_i^T \boldsymbol{\theta}])} f(\boldsymbol{\theta}) d\boldsymbol{\theta} \qquad (31)$$

We can recognise this particular latent variable representation of the Ising model as a multidimensional IRT model[47] from the common cause framework, where the vector **Θ** represent the set of latent abilities measured by the items in **X**. In addition to this vector, we also find the matrix *A* in the model, where the *i*[th] row contains the discrimination parameters for all latent variables on item *i*. In the traditional interpretation of the IRT framework, the discrimination parameter quantifies how well the item measures the corresponding latent variable, or in model terms, the degree to which the probability for item responses varies with respect to each latent variable in **Θ**. We obtain the following expression for the conditional probabilities of **X** given the vector of latent variables (**Θ**):

$$p(\mathbf{x}|\boldsymbol{\theta}) = \prod_i \frac{\exp(x_i[\delta_i + \boldsymbol{\alpha}_i^T \boldsymbol{\theta}])}{\sum_{\mathbf{x}} \exp(x_i[\delta_i + \boldsymbol{\alpha}_i^T \boldsymbol{\theta}])} \qquad (32)$$

Note that, as the number of non-zero eigenvalues represents the number of latent dimensions in the model, under the condition that only the first eigenvalue is non-zero, and the discrimination parameters with respect to the single resulting latent variable are 1 for all the items, the model reduces to the Rasch model from equation (1).

*Eigenvalue to Collider representation.* To acquire an collider representation of the Ising model we start again from the eigenvalue representation of the Ising model:

$$p(\mathbf{x}) = \frac{1}{Z} \exp\left( \sum_i x_i \delta_i + \sum_r \frac{1}{2} \lambda_r \left[ \sum_i q_{ir} x_i \right]^2 \right) \qquad (33)$$

By taking the partition function out of the expression we obtain the following proportionality relation:

$$p(\mathbf{x}) \propto \exp\left( \sum_i x_i \delta_i + \sum_r \frac{1}{2} \lambda_r \left[ \sum_i q_{ir} x_i \right]^2 \right) \qquad (34)$$

Next we can introduce a set of (independent) effect variables (**E** = [$E_1$, $E_2$, ..., $E_m$]) for each eigenvalue, such that we obtain a collider representation of the Ising model where endogenous selection bias has taken place for multiple effect variables. To that end, we recognise $p(\mathbf{x})$ from equation (34) as $p(\mathbf{x}|\mathbf{e}=1)$, the conditional probability of the collective cause (**X**) given that all effects are present (**E** = 1), proportional to the product of the marginal distribution for the collective cause $[p(\mathbf{x}) \propto \sum_i \exp(x_i \delta_i)]$, and the probability of observing the effects given the collective cause $\left[ p(\mathbf{e}=1|\mathbf{x}) \propto \exp\left( \sum_r \frac{1}{2} \lambda_r [\sum_i q_{ir} x_i]^2 \right) \right]$. Taking the sum over *i* in $p(\mathbf{x})$, and the sum over *r* in $p(\mathbf{e}=1|\mathbf{x})$ out of their respective exponential, and adding the appropriate normalising constant to make the probabilities sum to one we obtain the following expression:

$$p(\mathbf{x}|\mathbf{e} = 1) = \prod_i \frac{\exp(x_i \delta_i)}{\exp(+\delta_i) + \exp(-\delta_i)} \prod_r \frac{\exp\left(\frac{1}{2} \lambda_r [\sum_i q_{ir} x_i]^2\right)}{\sup_{\mathbf{x}} \exp\left(\frac{1}{2} \lambda_r [\sum_i q_{ir} x_i]^2\right)} \qquad (35)$$

Such that we can write the joint distribution of causes and effect variables as a common effect representation of the Ising model:

$$p(\mathbf{x}, \mathbf{e}) = \prod_i \frac{\exp(x_i \delta_i)}{\exp(+\delta_i) + \exp(-\delta_i)} \prod_r \left( \frac{\exp\left(\frac{1}{2} \lambda_r [\sum_i q_{ir} x_i]^2\right)}{\sup_{\mathbf{x}} \exp\left(\frac{1}{2} \lambda_r [\sum_i q_{ir} x_i]^2\right)} \right)^{e_r}$$

$$\times \left( 1 - \frac{\exp\left(\frac{1}{2} \lambda_r [\sum_i q_{ir} x_i]^2\right)}{\sup_{\mathbf{x}} \exp\left(\frac{1}{2} \lambda_r [\sum_i q_{ir} x_i]^2\right)} \right)^{1-e_r} \qquad (36)$$

We can quickly recognise a collider model in this distribution that is extended for as much common effect variables as there are non-negative eigenvalues. With this we have completed our second, and final, set of proofs, showing that three, statistically equivalent, representations of the Ising model exist that explain observed associations between binary variables as arising either through marginalisation with respect to latent variables, through reciprocal affect between variables, or through conditioning on common effect variables.

## Acknowledgements

We thank Dr. Maria A. Bolsinova and Dr. Maarten Marsman for proof reading the manuscript. This work was supported by by NWO (The Netherlands Organisation for Scientific Research), No. 022.005.0 (JK), and No. CI1-12-S037 (GM).

#### Author Contributions
J.K. and G.M. wrote the main manuscript. All authors reviewed the manuscript.

#### Additional Information
**Competing financial interests:** The authors declare no competing financial interests.

**How to cite this article**: Kruis, J. and Maris, G. Three representations of the Ising model. *Sci. Rep.* **6**, 34175; doi: 10.1038/srep34175 (2016).